\documentclass[12pt,a4paper]{article}
\usepackage{mathpazo}
\usepackage{amsfonts}
\usepackage[usenames]{color}
\usepackage{array}
\usepackage[intlimits]{amsmath}
\usepackage{mathrsfs}
\usepackage{booktabs}
\usepackage{fancyhdr}
\usepackage{epsfig}
\usepackage{verbatim}

\usepackage{graphicx}% Include figure files
\usepackage{dcolumn}% Align table columns on decimal point
\usepackage{bm}% bold math
\usepackage{epsf}

\newcommand{\beq}{\begin{equation}}
\newcommand{\eeq}{\end{equation}}
\newcommand\beqa{\begin{eqnarray}}
\newcommand\eeqa{\end{eqnarray}}
\newcommand\bea{\begin{array}}
\newcommand\eea{\end{array}}
\newcommand\ba{\begin{array}}
\newcommand\ea{\end{array}}
\newcommand{\nn}{\nonumber}
\newcommand{\diag}[1]{{\rm diag}(#1)}
\newcommand{\neqa}{\nonumber\end{eqnarray}}
\newcommand{\la}{\label}
\newcommand{\vrho}{{\varrho}}

\renewcommand{\P}{{\cal P}}

\newcommand{\eq}[1]{eq.(\ref{#1})}
\newcommand{\eqs}[2]{eqs.(\ref{#1},\ref{#2})}

\newcommand{\ur}[1]{(\ref{#1})}

\newcommand{\Tr}{{\rm Tr}}

\newcommand{\cv}{{\cal V}}
\newcommand{\Det}{{\rm Det}}
\newcommand{\half}{\frac{1}{2}}

\renewcommand{\d}{\partial}
\renewcommand{\O}{{\cal O}}

\renewcommand{\>}{{\rangle}}
\newcommand{\D}{r_{12}}

\newcommand{\cB}{{\cal B}}
\newcommand{\bv}{\overline{{\rm v}}}

\renewcommand{\)}{\right)}
\renewcommand{\(}{\left(}
\renewcommand{\]}{\right]}
\renewcommand{\[}{\left[}
\renewcommand{\v}{{\rm v}}

\newcommand{\re}{\relax{\rm I\kern-.18em R}}

\def\su2{{SU(2)}}

\def\tr{{\rm tr}}
\def\m{{\rm m}}
\def\s{{\rm s}}
\def\r{{\rm r}}

\begin{document}
\begin{center}
{\Large\bf Determinant of the SU(N) caloron with nontrivial holonomy}\\
\vspace{1cm}
%Nikolay Gromov\footnote{nikgromov@gmail.com}$^{a,b}$,\hspace{0.3cm}
Sergey Slizovskiy\footnote{Sergey.Slizovskiy@teorfys.uu.se}$^{a,b}$\\
\vspace{1cm}
{\it\footnotesize$^a$ St.Petersburg INP, Gatchina, 188 300, St.Petersburg, Russia \\ \vspace{.2cm}
%$^b$ Laboratoire de Physique Th\'eorique
%de l'Ecole Normale Sup\'erieure et l'Universit\'e Paris-VI,
%Paris, 75231, France\\ \vspace{.2cm}
$^b$ Department of Theoretical Physics,
Uppsala University,
P.O. Box 803, S-75108, Uppsala, Sweden}
\end{center}
\vspace{1cm}
\begin{abstract}
The 1-loop quantum weight of the SU(N) KvBLL caloron with nontrivial holonomy is calculated.
%for the case of well-separated constituent dyons.
The latter is the most general self-dual solution with unit topological charge
in the 4d Yang-Mills theory with one compactified dimension (finite temperature).
\end{abstract}

\section{Introduction}
 The finite temperature field theory is defined by considering the Euclidean
space-time which is compactified in the `time' direction whose inverse
circumference is a temperature $T$, with the usual periodic
boundary conditions for boson fields and anti--periodic conditions
for fermion fields. In particular, the gauge field
is periodic in time, so the theory is no longer invariant under
arbitrary gauge transformations. Only time-periodic gauge
transformations are allowed and hence the number of gauge invariants increases.
The new
invariant is the holonomy or the eigenvalues of the Polyakov line
that winds along the compact 'time' direction~\cite{Polyakov}:
\beq
L= \left.{\rm
P}\,\exp\left(\int_0^{1/T}\!dt\,A_4\right)\right|_{|\vec
x|\to\infty}. \la{Pol0}
\eeq\\
This invariant together with the topological charge and the magnetic
charge can be used for the classification of the field
configurations \cite{GPY} , its zero vacuum average is
one of the common criteria of confinement.

A generalization of the usual Belavin--Polyakov--Schwartz--Tyupkin
(BPST) instantons~\cite{BPST} for arbitrary temperatures and holonomies is the
Kraan--van Baal--Lee--Lu (KvBLL) caloron with non-trivial
holonomy~\cite{KvB,KvBSUN,LL}. It is a self-dual electrically
neutral configuration with unit topological charge and arbitrary
holonomy. This solution was constructed by Kraan and van Baal
\cite{KvB} and Lee and Lu \cite{LL} for the SU(2) gauge group and in
\cite{KvBSUN} for the general $SU(N)$ case; it has been named the
KvBLL caloron (recently the exact solutions of higher topological
charge were constructed and discussed \cite{KvBmult,KvBmult1}).
There is a plenty of lattice studies supporting the presence of these
solution \cite{IMMPSV}, see also \cite{vanBaal2006} for a very brief review. In a recent
paper \cite{Mitia} the caloron ensemble was studied analytically, although some contributions
were neglected there, the results are in very good agreement with phenomenology.

The holonomy is called 'trivial' if the Polyakov loop (\ref{Pol0})
acquires values belonging to the group center $Z(N)$.
For this case the KvBLL caloron reduces to the periodic Harrington-Shepard
\cite{HS} caloron known before. The latter is purely an $SU(2)$ configuration
and its quantum weight was studied in detail by Gross, Pisarski and Yaffe
\cite{GPY}.

The KvBLL caloron in the theory with $SU(N)$ gauge group on the
space $R^3\times S^1$  can be interpreted as a composite of $N$
distinct fundamental monopoles (dyons)~\cite{LY}\cite{Wein} (see
fig. \ref{fig_zt}
%and fig. \ref{fig_zx}
). As was proven in
\cite{KvBSUN,GSFermSUN},  the exact
KvBLL gauge field reduces to a superposition of BPS dyons, when the
separation $\vrho_i$ between dyons is large (in units of inverse
temperature). On the contrary, the KvBLL caloron reduces to the
usual BPST instanton, when the distances $\vrho_l$ between all the dyons
become small compared to the inverse non-triviality of holonomy.

  We refer the reader to the papers \cite{KvBSUN,GSFermSUN} for the detailed discussion and construction of the
caloron solutions, to the original works \cite{KvB} for the $SU(2)$ case and to further works on higher topological
charge solutions \cite{KvBmult,KvBmult1}.

  This paper is in the series of papers \cite{DGPS,GS,GSFermSUN,GromovAndrews,GromovSU2} where we calculate the functional determinant
for KvBLL calorons with nontrivial holonomy \cite{KvB,LL} in the finite-temperature Yang-Mills theory.

%  KvBLL calorons were extensively
%studied on the lattice in ~\cite{Brower,IMMPSV,Gatt} for $SU(2)$ and
%$SU(3)$ gauge groups.
%Up to now, only the determinants in case of the $SU(2)$ gauge group
%were found. In ref.~\cite{DGPS} the determinant for gluons and
%ghosts for the $SU(2)$ Yang--Mills theory was computed. It was
%extended to the SU(2) Yang-Mills theory with light fermions in
%\cite{GS}. So far only a metric of the moduli space was known for
%the general $SU(N)$ case \cite{Kraan} (its determinant was analyzed in
%details in \cite{DG}). The fermionic zero-modes were studied in
%\cite{Cherndb}.

%%%%%%%%%%%%%%%%%%%%%%%%%%%%%%%%
% FIGURE 1
%%%%%%%%%%%%%%%%%%%%%%%%%%%%%%%%
\begin{figure}[t]
\centerline{
\epsfxsize=0.4\textwidth
\includegraphics[height=\epsfxsize,angle=-90]{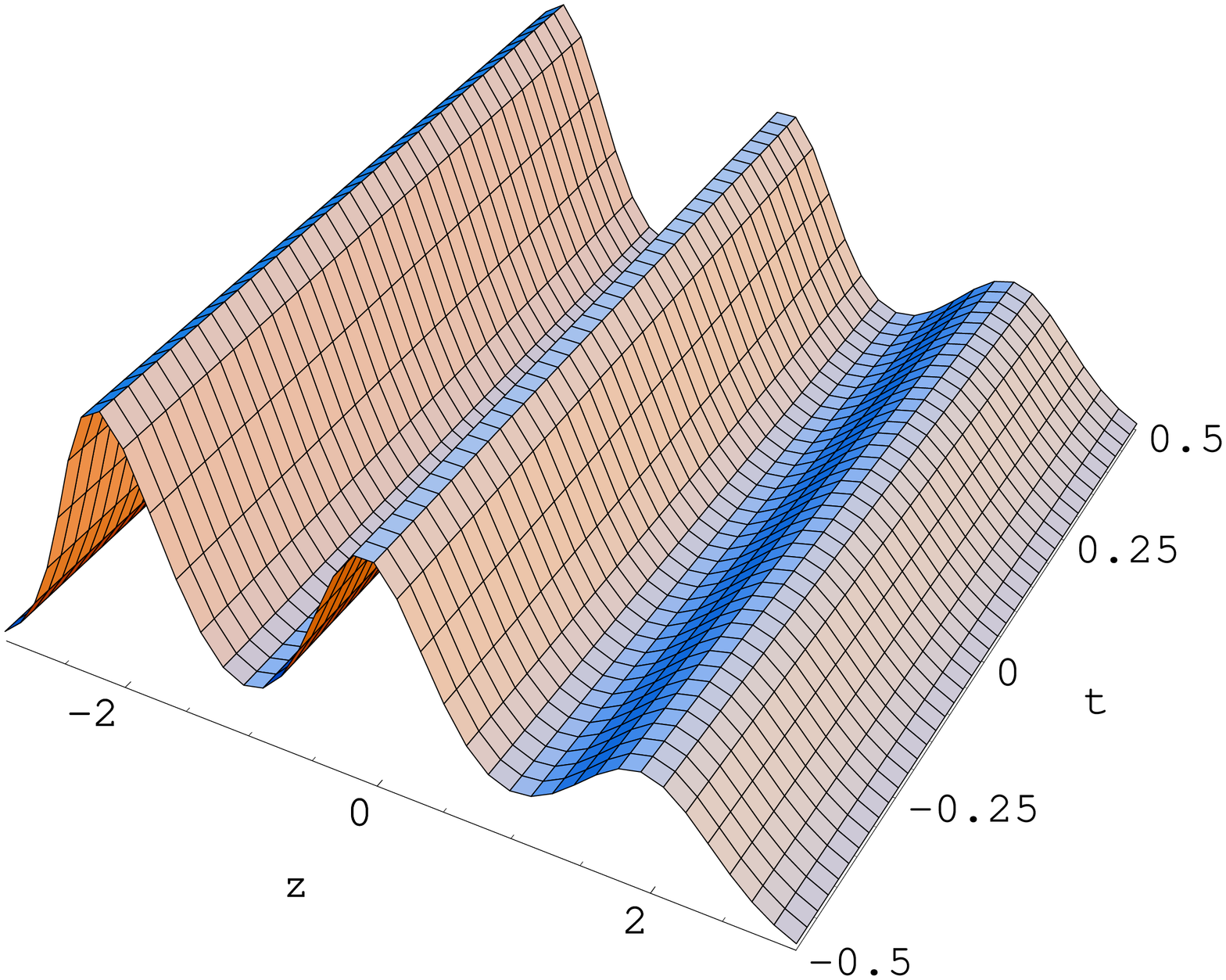}
\epsfxsize=0.4\textwidth
\includegraphics[height=\epsfxsize,angle=-90]{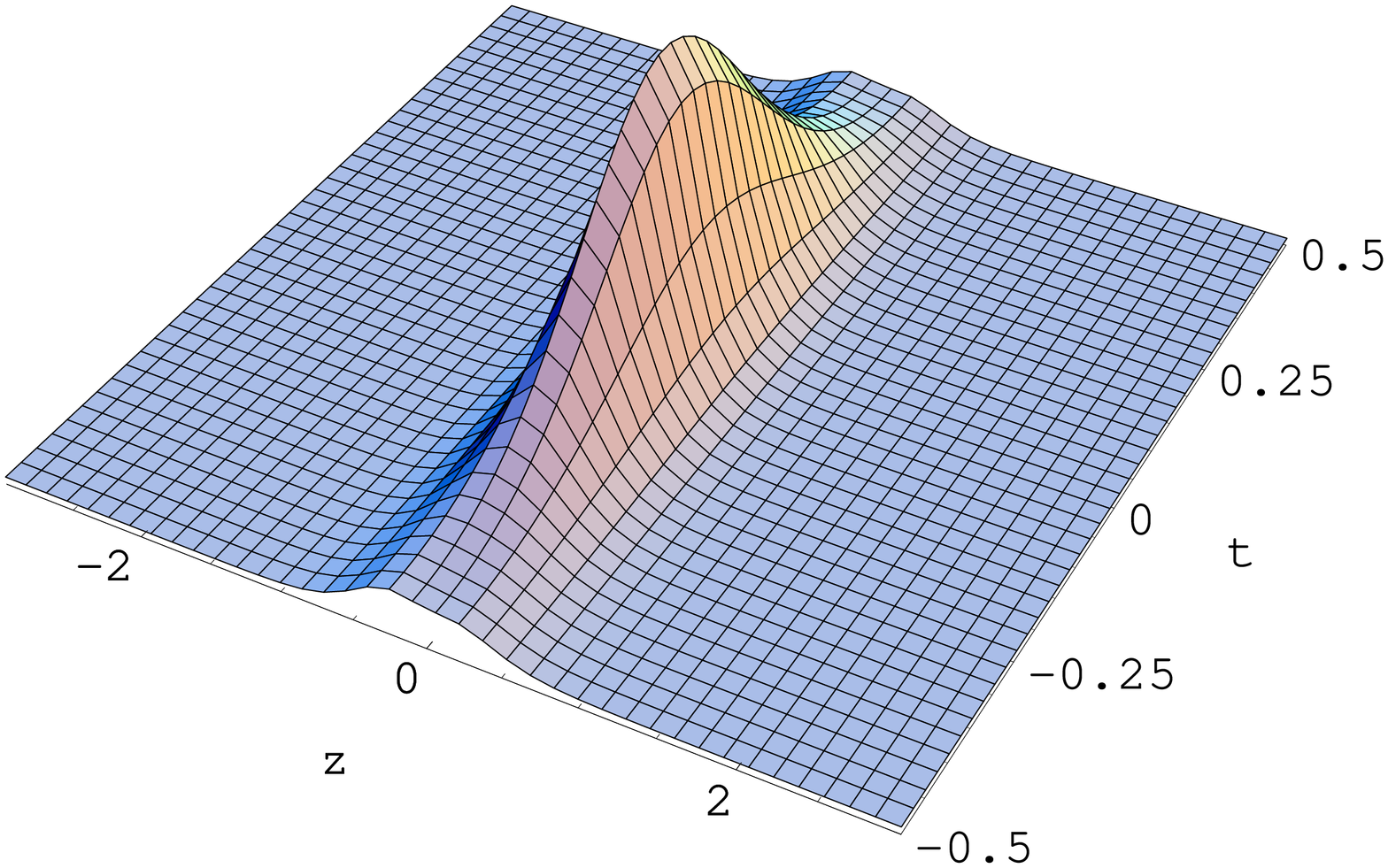}}
\caption{The action density of the $SU(3)$ KvBLL caloron as function
of $z,t$ at fixed $x=y=0$, eigenvalues of $A_4$ at spatial
infinity are $\mu_1=-0.307 T,\;\mu_2=-0.013 T,\;\mu_3=0.32 T$. It is periodic in $t$  direction.
At large dyon separation the density becomes static (left, $\vrho_{1,2}=1/T,\;\vrho_3=2/T$).
As the separation decreases the action density becomes more like a $4d$
lump (right, $\vrho_{1,2}=1/(3T),\;\vrho_3=2/(3T)$).
The axes are in units of inverse temperature $1/T$.\la{fig_zt}}
\end{figure}
%%%%%%%%%%%%%%%%%%%%%%%%%%%%%%%%

 Here we calculate the 1-loop gluonic and ghost functional determinants
for the case of an arbitrary $SU(N)$ gauge group.  The calculation is performed in the
limit of far separated
dyon constituents and up to an overall numerical constant. The constant for the
gluonic determinant remains known only for the $SU(2)$ case \cite{DGPS}
\footnote{In the previous paper \cite{GSFermSUN}
we have proved
%by a lengthy computation
that the corresponding constant is zero for the
fundamental-representation determinant, but for the adjoint
representation the constant is not expected to vanish.}.

We find new 3-particle interactions arising between constituent dyons due to the gluonic determinant.
These terms
were not present in the fermionic (fundamental representation) determinant and also vanished in
the $SU(2)$ gluonic determianant.

%The reader not interested in the
%calculation itself to the Section \ref{SecFinalResult} where the final result is presented.
 Similar to our previous results, the determinant
is infrared divergent, the leading divergence is proportional to
the volume or the system, and there are of course several
subleading divergent terms. It is not surprising and was known
long ago \cite{NW, GPY} that non-trivial holonomy increases the
effective action by a factor proportional to the volume. Nevertheless that does
not make the studies of non-trivial holonomy unphysical, since in
the ensemble of many calorons the moduli space integrals can
compensate the above divergences.

Since there are subleading divergences coming from the Coulomb tail of dyon fields, it is natural, that
the result would also depend on the position of the large ball, with which we make the infra-red cutoff.
We will display this dependence, but we note that it is unphysical unless the box in not a real border
of plasma region.
%in the sense that in the ensemble
%of many calorons one has to remove the IR-cutoff ball and add interaction terms with other dyons instead.
One could also expect that the quantum corrections will dump the Coulomb tails of dyons.
% lead to the same type of divergence with opposite sign.

We present the relevant notations in section \ref{secNotations} and illustrate the
notations by the old results.  The method of computation
is described in \ref{secMethod} and the actual computations are carried out in
subsequent sections and appendices.
The final result is presented in section \ref{SecFinalResult}.

We do not draw here any physical conclusions on the behaviour of the whole
caloron ensemble since that is now a separate business \cite{GasModel}, \cite{Mitia}.
Our results
could be useful for extending the works \cite{GasModel} to the $SU(N>2)$ case and including
the corrections due to the non-zero modes to the work \cite{Mitia}.
% We draw attention of the reader
%to the work \cite{Mitia}, presenting analytical results on caloron ensemble that strikingly well match the phenomenology
%even at zero temperature.

\section{Notations and Review} \la{secNotations}
Consider the $SU(N)$ YM theory and a caloron solution with the
asymptotics \footnote{We use notations consistent with \cite{KvBSUN}.}
$$A_\mu \longrightarrow^{\!\!\!\!\!\!\!\!\!\!\!\!\vec x \to \infty} 2 \pi \delta_{\mu4}\, \diag{\mu_1,...,\mu_N}.$$

For the $SU(2)$ case the standard choice is $\mu_1 = - \omega \ ; \mu_2 = \omega$ where
$0 \le \omega <\frac12$.
As usual, we set the temperature $T=1$ throughout the computation,
and restore the temperature dependence only in the final result.

The caloron can be viewed as composed of dyons
(BPS monopoles with $A_4$ playing the role of a Higgs field),
 the inverse dyon size $\nu$ being defined as
\beq
\la{defnu}
\nu_l = \mu_{l+1}-\mu_{l} \;\;;\;\; \nu_N = \mu_1-\mu_N+1.
\eeq
Traditionally the first $N-1$ dyons are called the 'M - dyons' and the $N^{-th}$ dyon
is called the 'L - dyon', because an additional gauge transformation
%$e^{2 \pi i x_4 \alpha_N \cdot t}
is need for it to
have correct asymptotics.

We also introduce a notation
\beq
\v_{mn}\equiv2\pi(\mu_m-\mu_n)\ {\rm mod}\ 2\pi.
\eeq
which coincides with $\v = \v_{21}$ and $\bar \v = \v_{12}$ used previously in the $SU(2)$ calculations.

The positions of dyon centers are denoted by $y_i$.
The distance from the $i^{-th}$ dyon center to a point $x$
is denoted as $\vec x - \vec y_i = r_i$; for the $SU(2)$ case the standard
notation is $r_1 = s \ ; \ r_2 = r$ \cite{DGPS}.
The distance between dyon cores is denoted by $r_{ij} = |\vec r_j - \vec r_i|$.

It is convenient to use a so-called 'algebraic gauge', in which the asymptotic gauge field
is vanishing at the expense of introducing twisted boundary conditions for field fluctuations.
%in the Euclidean time direction.
The twist $a(\vec x,1/T)=e^{-i\tau}a(\vec x,0)$ is hence related to the holonomy as
$\tau = 2 \pi \diag{\mu_1, ... ,\mu_N}$.  The holonomy and, correspondingly, the twist
could also be multiplied by elements
of the center of the gauge group $e^{2 \pi i \frac{ k}{N}}$. It does not affect the adjoint gauge
field and determinant but it affects fundamental determinants \cite{GSFermSUN}:
\beqa
\nn\log\det(-\nabla^2_N)&=&\sum_n\(\frac{\pi}{4}P''(\tau_n)r_{n,n-1}+\frac{1}{2}P(\tau_n)V^{(3)}
-\frac{\nu_n\log\nu_n}{6}-\frac{\log r_{n,n-1}}{12\pi r_{n,n-1}}\)\\
&&+c_N+\frac{1}{6}\log\mu+\O(1/r)
\eeqa
where
\beq
c_N=-\frac{13}{72}-\frac{\pi^2}{216}+\frac{\log\pi}{6}-\frac{\zeta'(2)}{\pi^2}.
\eeq
$P$ is a periodical function with a period $2\pi$ such that
\beq
P(\v)=\frac{q^2(2\pi-q)^2}{12\pi^2}; \;\; q=\v \!\! {\rm mod} 2 \pi
\eeq

%Determinant in fundamental representation $SU(N)$ is

Determinant in the adjoint representation of $SU(2)$ reads \cite{DGPS,GromovSU2}
\beqa
&&\log\Det(-D^2_2)=V P(\v)+ 2\pi P''(\v)\D
+\frac{3\pi-4\v}{3\pi}\log\v
+\frac{3\pi-4\bv}{3\pi}\log\bv
\nn \\\nn && +\frac{2}{3}\log\mu +\frac{5}{3}\log(2\pi)+c_2
\\
&&
+\frac{1}{\D}\left[\frac{1}{\v}+\frac{1}{\bv}
+\frac{23\pi}{54}-\frac{8\gamma_E}{3\pi}-\frac{74}{9\pi}
-\frac{4}{3\pi}\log\left(\frac{\v\bar\v\,\D^2}{\pi^2}\right)\right]
+\O\left(\frac{1}{\D^2}\right)
\eeqa

Now we proceed to the calculation of $SU(N)$ adjoint-representation determinant.

\section{Method of computation} \la{secMethod}
For self-dual fields the gluonic and ghost determinants over non-zero modes for the background gauge fixing
are related \cite{BC} to the
adjoint scalar determinant in the same background: $\Det'(W_{\mu\nu})= \Det(-D^2)^4$, where  $W_{\mu\nu}$
is the quadratic form for spin-1, adjoint representation quantum fluctuations and  $D^2$ is
the covariant Laplace operator for spin-0, adjoint representation ghost fields. So the total contribution to the effective
action of gluon and ghost determinants is $2 \log \Det(-D^2)$ which corresponds to two physical degrees of freedom.

We calculate the quantum determinant by integrating its variation with respect
to parameters ${\cal{P}}$ of the solution, following
\cite{GSFermSUN,GS,DGPS,Zar}. In this case the problem reduces to four dimentional
integral of the gauge field variation multiplied by a vacuum current, which can be
expressed through Green function known implicitly for any self-dual configuration
\beq \frac{\partial\,\log {\rm
Det}(-D^2[A])}{\partial {\cal P}} = \!-\!\int
d^4x\,\Tr\left(\partial_{\cal P} A_\mu\, J_\mu\right)
\label{dvDet}
\eeq
where $J_\mu$ is the vacuum current in the
external background,  determined by the Green function:
\beq
 J_\mu\equiv \overrightarrow{D}_\mu {\cal G}
+{\cal G} \overleftarrow{D}_\mu. \label{defJ}
\eeq
Here $\cal G$
is the periodical Green function of the covariant Laplas operator in adjoint representation
\beqa
-D^2_x{ G}(x,y)&=& \delta^{(4)}(x-y) \la{Gdef} \\
{\cal G}(x,y)&=& \sum_{n= -\infty}^{+\infty} G(x_4,{\vec
x};y_4+n,{\vec y}).
\eeqa
The Green functions in the self-dual backgrounds are
known explicitly \cite{CWS,Nahm80} if the gauge field
is expressed in terms of
the Atiyah--Drinfeld--Hitchin--Manin (ADHM) construction~\cite{ADHM}: $A_\mu=v^\dag\d_\mu v$.
%(we refer the reader
%to \cite{GSFermSUN} and to Section \ref{secRegCurrent}  for a review of ADHMN construction).
These look quite
simple for the fundamental representation \cite{CorriganGreen}
\beq
\label{green12} G^{\rm fund}(x,y)= \frac{ v^\dag(x)v(y)}
{4\pi^2(x-y)^2}\;,
\eeq
but become more complicated for the adjoint representation \cite{CorriganTensor,NahmGreen,Jack}:
\beqa\label{green1}
G^{ab}(x,y)&= &\frac{\half\Tr\,t^a\langle v(x)|v(y)\rangle t^b
\langle v(y)|v(x)\rangle} {4\pi^2(x-y)^2}  \nonumber\\
&+&\frac{1}{4\pi^2}\int_{-1/2}^{1/2}dz_1\,dz_2\,dz_3\,dz_4\, M(z_1,z_2,z_3,z_4)\\
\nn&\times&\half\Tr\!\left(\cv^{\dagger}(x,z_1) \cv(x,z_2) t^a\right)\Tr\!\left(\cv^{\dagger}(y,z_4)\cv(y,z_3) t^b\right)\,,
\eeqa
where $t^a$ are Hermitian fundamental-representation generators of $SU(N)$ normalized to
$\tr\, t^a t^b = \frac12 \delta ^{a b}\ $;  $\cv(x,z)$ is one of the components of $v$ (see \eq{defv12})
%\beq
%M_{pqnm}= \int_{-1/2}^{1/2}M(z_1,z_2,z_3,z_4)\,
%e^{2\pi i (\!-\!p z_1\!+\!q z_2\!+\!n z_3\!-\!m z_4)}
%dz_1dz_2dz_3dz_4.
%\eeq
and $M$ is a piece-wise rational function\footnote{see Appendix B to \cite{DGPS}
for its explicit form in case of $SU(2)$}. Fortunately we do not need
an explicit form of this function for the $SU(N)$ caloron since
in the large separation limit the contribution of the last term (or ``M-term'')
is exponentially small away from the dyons. Near the dyons the field
is essentially reduced to the $SU(2)$, so one can use there the results of \cite{NahmGreen, DGPS}.

In what follows it will be convenient to split the periodic propagator  into
three parts and consider them separately:
\beqa \nn
{\cal G}(x,y)&=& {\cal G}^\r(x,y)+{\cal G}^\s(x,y) + {\cal G}^\m(x,y) ,\\
\la{green3}
&&{\cal G}^\s(x,y)\equiv G(x,y),
\\&&{\cal G}^\r(x,y) + {\cal G}^\m(x,y)
\nn\equiv\sum\limits_{n\neq 0}  G(x_4,{\vec x};y_4+n,{\vec
y})\la{GrGs}\;,
\eeqa
here ${\cal G}^\m(x,y)$ coresponds to the part of the propagator arising from the $M$-term.
The vacuum current \ur{defJ} will be also split into
three parts, ``singular'' , ``regular'' and ``M'', in accordance with (\ref{green3})

\beq J_\mu= J^\r_\mu+J^\s_\mu+ J^\m_\mu. \eeq

As was proposed in \cite{DGPS} we divide the space into regions surrounding the dyons
 and the remaining space (outer region). Near each of the dyons the gauge field
 becomes essentially the $SU(2)$ dyon configuration
% plus a correction of order $\O(1/r_{ij})$ and
plus an additional constant-field background.
In this region we can use the results of \cite{DGPS}. In the outer region,
far from the exponential cores of the dyons, the vacuum current considerably simplifies and we only have to
perform integration in (\ref{dvDet}).

In the following two sections we give results for these two domains and
in section \ref{secIntegration} we combine them together and integrate over the space.

\section{Core domain}
In this section we  write a contribution to
the variation of total determinant arising from the core region of a dyon.
We take a ball of radius $R$ around the dyon.
In that region the field is approximately the one of the $SU(2)$-dyon, embedded along one of the simple roots, {\it plus}
an extra constant  $A_4$ field \cite{KvBSUN}.
More precisely in the fundamental representation the gauge field near the $l^{\rm th}$ dyon
is a zero $N\times N$ matrix with only
$2\times 2$ block at $l^{\mbox{-th}}$ position filled by the BPS dyon gauge field,
plus a constant diagonal $N\times N$ matrix \cite{GSFermSUN}
\beqa\la{Amunear}
&&A^{ l^{\rm th}{\rm\; block}\;2\times 2}_\mu=A^{\rm dyon}_\mu(\nu_l,\vec x-\vec y_l)+2\pi i\left(\frac{\mu_l+\mu_{l+1}}{2}\right) \delta_{\mu4} 1_{2\times2}\;,\\
\nn &&A^{{\rm outside}\;l^{\rm th}{\rm\; block}\;2\times 2}_\mu=2\pi i\;{\rm diag}\left(\mu_1,\dots,\mu_N\right) \delta_{\mu4}
.\eeqa
Under the action of the $SU(2)$ sub-group, the adjoint representation of $SU(N)$ splits
into one triplet, $2(N-2)$ doublets and $(N-2)^2$ singlets.
The determinant of arbitrary $SU(2)$ configuration embedded into $SU(N)$
is then expressed as a sum of $SU(2)$ adjoint-representation
determinant plus $2(N-2)$ fundamental-representation determinants
\footnote{See for example \cite{Bernard77} where that was done for the instanton solution.}.

As is seen from \eq{Amunear}, the $SU(2)$ dyon field is accomplished by the constant diagonal
matrix. This matrix can be killed by a gauge transformation, which is not periodical,
and thus can change the determinant. It is equivalent to the additional twist of boundary
conditions for the
$2(N-2)$ fundamental representation determinants.

As a demonstration let us consider the $SU(3)$ case. The fundamental gauge field reads
\beq
A_\mu=
\(\bea{c|c}
A_{\mu\ 2\times 2}^{dyon} & \bea{c}0\\0\eea\\ \hline %\midrule
\bea{cc}0 & 0\eea & 0
\eea\)+2\pi i\delta_{\mu 4}
\(\bea{ccc}
\frac{\mu_1+\mu_2}{2} & 0 & 0\\
0 & \frac{\mu_1+\mu_2}{2} & 0\\
0 & 0 & \mu_3
\eea\)
\eeq
in the adjoint representation of $SU(3)$
in an appropriate basis it becomes
{\small
\beqa
\hat A_\mu=\left(
\bea{c|c|c|c}
\hat A_{\mu\ {3\times 3}}^{dyon}
& 0 & 0 &0 \\ \hline %\midrule
0 &
-A_{\mu \ {2\times 2}}^{dyon}-i\pi \delta_{\mu4}(\nu_3-\nu_2) & 0 & 0 \\ \hline %\midrule
0 &0 &
A_{\mu \ {2\times 2}}^{dyon}+i\pi \delta_{\mu4}(\nu_3-\nu_2) & 0 \\ \hline %\midrule
0 & 0 & 0 & 0
\eea\right)
\eeqa}

So that there is one block $3\times3$ giving adjoint representation dyon field and two  $2\times 2$
blocks, giving fundamental representation dyon accomplished by a unit matrix. As it was shown
in \cite{GSFermSUN} this extra unit matrix changes only an IR divergent part of the dyon determinant (the one,
 depending on radius of the ball).
These divergences cancel with the terms in the outer-region
determinant depending on the radius of the holes $R$,   as it is shown in
Appendix \ref{AppIR}).
So we can freely drop them.

Summing up one adjoint (\ref{intM2}) and $2(N-2)$ fundamental dyon determinants (\ref{intM}) for all $N$ dyons we obtain
the following contribution to the derivative of the Caloron determinant from the considered domain
\beq\la{DetCore}
\d_\P \sum_n\left( -\frac{(6+N)\nu_n\log(\nu_n)}{3}+\log(\nu_n) \right)+{\rm IR},
\eeq
where ``IR" denotes the IR divergent terms.

\section{Outer domain} \la{secFar}
We proceed to consider the far domain, i.e. the region of space outside dyons' cores.
Caloron field becomes diagonal with $\O(e^{-\nu_i r_i})$ precision and this
simplifies significantly the results. For instance the $4^{th}$ component of the fundamental caloron gauge field reads
\beq
A_4^{mn}=i\delta^{mn}\(2\pi\mu_m+\frac{1}{2r_m}-\frac{1}{2r_{m-1}}\)
\la{A4ep}
\eeq
In what follows we will consider the derivative of the determinant with respect to $\mu_m$. It turns out
that in this domain only $A_4$ depends on $\mu_m$ nontrivially $\cite{KvBSUN}$. Thus we need only
$4^{th}$ component of the vacuum current as it follows from (\ref{dvDet}).
As we know from the $SU(2)$ case this component of the current is especially simple
\cite{DGPS}
\beq
J_4^{\rm su(2)}=\frac{i}{2}T_3 P'\(v+\frac{1}{r_1}-\frac{1}{r_2}\)
\eeq
the natural generalization of this expression is
\beq
J_4^{\rm su(n)}=\!\!\!\bea{c} _{N} \\{\rm diag} \\ ^{ n,m=1} \eea\!\!\!\[\frac{i}{2}P'\(2\pi(\mu_m-\mu_n)+\frac{1}{2r_m}-\frac{1}{2r_{m-1}}-\frac{1}{2r_n}+\frac{1}{2r_{n-1}}\)\]
\eeq

The expression in the brackets is simply the eigenvalue of the
the gauge field (\ref{A4ep}) in the adjoint representation. This formula is definitely
right for large $r_m$, where the field becomes almost constant \cite{GPY,DO} and generalizes the $SU(2)$
expression. Moreover we check it by a direct computation in Appendix \ref{AppA}. We conclude
that
\beq\la{toint}
-\tr(\d_{\cal P}A_\mu J_\mu)\!=\!\frac{1}{2}\sum_{n,m}\d_{\cal P}
P\(\v_{mn}+\frac{1}{2r_m}-\frac{1}{2r_{m-1}}-\frac{1}{2r_n}+\frac{1}{2r_{n-1}}\!\)
\eeq
where
\beq
\v_{mn}\equiv2\pi(\mu_m-\mu_n)\
%{\rm mod}\ 2\pi.
\eeq
The variation over ${\cal P}$ can be integrated up to a constant, and the integral over space would be performed
in the next section.

\subsection{Integration \la{secIntegration}}
In order to get a variation of the determinant we have to integrate in \eq{toint} over the space with $N$ spherical holes of radius R .
The following integrals will be very helpful
\beqa
\la{integr1}
\nn\lefteqn{\int\(\frac{1}{2r_m}-\frac{1}{2r_{m-1}}-\frac{1}{2r_n}+\frac{1}{2r_{n-1}}\)^2d^3 x\simeq}\\
\la{int2}&&\pi\( r_{m,n}+r_{m-1,n-1}-r_{m,n-1}-r_{m-1,n}+r_{m,m-1}+r_{n,n-1}\)\\
\nn\lefteqn{\int\(\frac{1}{2r_m}-\frac{1}{2r_{m-1}}-\frac{1}{2r_n}+\frac{1}{2r_{n-1}}\)^3d^3 x\simeq}\\
\la{int3}&&-3\pi\log\(\frac{r_{m,n-1}}{r_{n,m-1}}\frac{r_{m,n,m-1}}{r_{n,m,n-1}}\frac{r_{n,m-1,n-1}}{r_{m,n-1,m-1}}\)
\eeqa
where $r_{nm}=|\vec y_n-\vec y_m|$ is a distance between dyons and
\beq
2 \, r_{lmn}\equiv r_{lm}+r_{mn}+r_{nl}
\eeq is
the perimeter of the triangle, formed by $l^{{\rm -th}},m^{{\rm -th}},n^{{\rm -th}}$ dyons.
Sign $\simeq$ means that we drop all the terms dependent on the radius of the holes $R$ since they cancels precisely
with dyons IR divergences as discussed in Appendix \ref{AppIR}.  To derive the last equation we used
\beq \nn\int \frac 1 {r_n r_m r_l}\ \ d^3x \simeq - 4 \pi \log r_{nml} + C .
\eeq
%Note that \eq{int2} is anti-symmetric.

 It is important to point out that the \eq{int3} is {\it not} applicable for the case $m=n \pm 1$, since it diverges.
 The reason
 is that the divergences near dyon cores are not balanced anymore. Nevertheless it is straightforward to verify
 that if one replaces a zero $r_{nn}$ under logarithm in \eq{int3} by some fixed $\epsilon$, then it is
 is still valid up to a constant, which
 cancels in the final result.

So we can integrate in \eq{toint}
\beqa\la{DetFar0}
\lefteqn{\int\frac{1}{2}\sum_{m,n}
P\(\v_{mn}+\frac{1}{2r_m}-\frac{1}{2r_{m-1}}-\frac{1}{2r_n}+\frac{1}{2r_{n-1}}\!\)d^3x\simeq}\\
\nn&&\sum_{m,n}\frac{\pi}{4}P''(\v_{mn})\( r_{m,n}+r_{m-1,n-1}-r_{m,n-1}-r_{m-1,n}+r_{m,m-1}+r_{n,n-1}\)\\
\nn&&-\sum_{m,n}\frac{[\v_{mn}]-\pi}{2\pi}\log\(\frac{r_{m,n-1}}{r_{n,m-1}}\frac{r_{m,n,m-1}}{r_{n,m,n-1}}\frac{r_{n,m-1,n-1}}{r_{m,n-1,m-1}}\)
+\sum_{n,m}\frac{1}{2}P(\v_{mn})V^{(3)}
\eeqa
We denote $[\v_{mn}]=\v_{mn}\ {\rm mod}\ 2\pi$.
To simplify the above expression we use the identity:
\beq
\sum_{m,n}\frac{[\v_{mn}]-\pi}{4\pi}\log\frac{r_{m,n,m-1}r_{n,m-1,n-1}}{r_{n,m,n-1}r_{m,n-1,m-1}}=
\sum_{m,n}\nu_n\log r_{m,n,m-1}-\sum_n\log r_{n,n-1}.
\eeq
Then \eq{DetFar0} becomes
\beqa\la{DetFar}
\lefteqn{{\log\det}(-D^2_N)^{far}=\int\frac{1}{2}\sum_{m,n}
P\(\v_{mn}+\frac{1}{2r_m}-\frac{1}{2r_{m-1}}-\frac{1}{2r_n}+\frac{1}{2r_{n-1}}\!\)d^3x\simeq}\\
\nn&&-\sum_{m,n}\frac{[\v_{mn}]-\pi}{2\pi}\log\(\frac{r_{m,n-1}}{r_{n,m-1}}\)-\sum_{m,n}2\nu_n\log r_{m,n,m-1}+2\sum_n\log r_{n,n-1}+\\
\nn&&\sum_{m,n}\frac{\pi}{4}P''(\v_{mn})\( r_{m,n}+r_{m-1,n-1}-r_{m,n-1}-r_{m-1,n}+r_{m,m-1}+r_{n,n-1}\)+\\
\nn&&+\sum_{m,n}\frac{1}{2}P(\v_{mn})V^{(3)}
\eeqa

The ``R-terms'' are exactly the ones of \eq{Rterms} but with $R$ standing as a lower limit of integration, this
provides the cancellation of them when we add the core contribution.
The second equality in \ur{DetFar} is valid when the variation
does not involve changing of the far region. Note that the $\frac{\log r}{r}$ correction comes only from
the far region. So we can calculate it. It comes from the next $P''''$ term in the Taylor series,
this term obviously involves 4-center Coulomb integrals:
\beq
 \int \frac{d^3 x}{r_1 r_2 r_3 r_4}
\eeq
taken over $\rm R^3$ with holes around centers. Since this integral converges both in the IR and UV (near the holes)
, it can involve only
logarithms of some dimensionless combinations of distances between these four points, divided by the distance.
In the approximation
that the dyons are spread homogeneously, such terms would be of order of unity and we neglect them. The
only {\it large} logarithms come from the case where three of the four points coincide,
in this case the integral diverges as logarithm near the $i$-th dyon:
\beq
\la{biglog}
 \int_{{\rm R^4} \setminus B_R} \frac1{r_i^3 r_j} = 4 \pi \frac {\log(r_{ij}/R)}{r_{ij}} + \O(1/r_{ij})
\eeq

So for the correction to $\log \Det (\!-\! D^2)$ one sums all the contributions of the form (\ref{biglog}).
Note that $P^{\rm IV} = \frac 2 {\pi}$ is a constant, so some terms cancel in the sum. The
result for $N>2$ is:
\beq \la{logcorr}
 \log \Det (\!-\! D^2) _{correction} = - \frac{6+N}{6 \pi} \sum_{n>m}^N \frac{\log r_{nm}}{r_{nm}},
\eeq
for $N=2$ the coefficient is doubled and becomes $-\frac8{3 \pi}$, since there are more coincident points.
It matches our $SU(2)$ result (eq. (60) in \cite{DGPS} ).

\section{The result} \la{SecFinalResult}
From \eqs{DetCore}{DetFar} we can conclude that for large dyons' separations, $\vrho_m\ll 1/\nu_m+1/\nu_{m-1}$,
the $SU(N)$ caloron determinant is the sum of these expressions plus some integration constant and $\frac{\log r}{r}$
improvement (\ref{logcorr}). Restoring the temperature dependence we obtain
\beqa\la{Det1}
&&{\log\det}(-D^2_N)=\\
\nn&&-\sum_{m,n}\frac{[\v_{mn}]-\pi}{2\pi}\log\(\frac{r_{m,n-1}}{r_{n,m-1}}\)-\sum_{m,n}2\nu_n\log r_{m,n,m-1}+2\sum_n\log r_{n,n-1}\\
\nn&&+\sum_{m,n}\frac{\pi}{4}P''(\v_{mn})T \( r_{m,n}+r_{m-1,n-1}-r_{m,n-1}-r_{m-1,n}+r_{m,m-1}+r_{n,n-1}\)\\
\nn&&+\sum_{m,n}\frac{1}{2}P(\v_{mn})T^3 V^{(3)}-\sum_n\frac{(6+N)\nu_n\log\nu_n}{3}+\sum_n\log\nu_n- \frac{6+N}{6 \pi} \frac{\log r_{nm} T}{r_{nm} T}+c_N
\eeqa
Note that the coefficient $- \frac{6+N}{6 \pi}$ should be doubled for $N=2$ case.

The contribution to the effective action from non-zero modes of gluons and ghosts would be
\beq
\delta S_{eff} =
-\log \frac{\Det(-D^2)}{  (\Det^{'}W_{\mu\nu})^{1/2}}  =  \log \Det( \!-\!D^2)
\eeq

The constant $c_N$ will of course contain a standard UV-divergence $c_N = c+ \frac{N}{3} \log{\mu_{PV}}$, coming
from the instanton determiant \cite{tHooft}, where $\mu_{PV}$
is a Pauli-Villars mass. This divergence, together with
$\left(\frac{\mu_{PV}}{g\sqrt{2\pi}}\right)^{4 N}$ coming from zero modes, gives the standard
Yang-Mills $\beta$-function and is commonly incorporated into the running coupling:
\beq
\mu_{PV}^{\frac{11}{3} N}\,e^{-\frac{8\pi^2}{g^2(\mu_{PV})}}= \Lambda^{\frac{11}{3} N}
\la{transmutation}
\eeq
where $\Lambda$ is the scale parameter obtained here through the
`transmutation of dimensions'.

Now let us combine with the result for $SU(N)$ caloron zero modes \cite{DG, Kraan} and the classical action $8 \pi^2/g^2(\mu_{PV})$.
The caloron measure is \cite{DG}
\beq
\int_{\cal G}\omega\simeq 2^{6N}\pi^{4N}\left[1+\sum_m\frac{1}{4\pi\vrho_m}
\left(\frac{1}{\nu_{m-1}}+\frac{1}{\nu_m}\right)\right]\prod_n\nu_n\;
d^3\vrho_1\dots d^3\vrho_{N-1}\; d^4\xi.
\la{SUN}\eeq

So the total contribution
of one caloron to the effective action becomes
\beqa
 e^{-S_{eff}} & \approx & \left(\frac{\Lambda e^{\gamma_E}}{4\pi T}\right)^{\frac{11}{3} N} \!\!\! C_N
 \int (\Det(\!-\!D^2_N))^{-1}  \left(\frac{8 \pi^{2}}{g^2(\mu_{PV})} \right)^{2 N} \times
 \nn\\&&\times \left[1+\sum_m\frac{1}{4\pi\vrho_m}
\left(\frac{1}{\nu_{m-1}}+\frac{1}{\nu_m}\right)\right]\prod_n\nu_n\;
d^3\vrho_1\dots d^3\vrho_{N-1}\; d^4\xi
\la{Zfull}
\eeqa
We have collected the factor $4\pi e^{-\gamma_E}T/\Lambda$ because it is  the natural
argument of the running coupling constant at nonzero  temperatures~\cite{coupling,DO}.
When we have done so in the $SU(2)$ case \cite{DGPS}, we have got a constant numerically very close to 1,
%so we expect that if we would manage to write the final result in a similar form,
so we expect
the constant $C$ to be of order of unity.

\section*{Acknowledgements}
I thank Nikolay Gromov for collaboration, discussions and editing this work.  We are grateful to
 Dmitri Diakonov, Victor Petrov, Konstantin Zarembo
and Michael Muller-Preussker for discussions.
%We also thank Dmitri Diakonov for critical reading of the manuscript.
This work was partially supported by
RSGSS-1124.2003.2 and by RFFI project grant 06-02-16786, the Dmitri Zimin 'Dynasty' foundation,
STINT Institutional Grant and by a Grant from VR.

\appendix
\section{Cancellation of IR divergenses of dyons}\la{AppIR}
Consider the field near the dyon constituent of $SU(N)$ caloron. In the fundamental representation it is given by the
\eq{Amunear}.  In the adjoint representation this field looks like one block with the $SU(2)$ BPS dyon in
the $SU(2)$-adjoint representation ($3 \times 3$)  and  $2 (N-2)$ blocks with the $SU(2)$ BPS dyon in
the $SU(2)$-fundamental representation ($2 \times 2$) {\it plus} a constant part, specified below:
\beqa
\la{DiagAdj}
  -i A_{const}^{adj} &=& \nn  \pi \, \diag{0,0,0,-(\mu_1+\mu_2 - 2\mu_3), -(\mu_1+\mu_2 - 2\mu_3), \\
  &&  \nn+(..),+(..), ... ,+(\mu_1+\mu_2 - 2\mu_N), +(\mu_1+\mu_2 - 2\mu_N),\\
  &&,\mu_3-\mu_4, \mu_4-\mu_3,...(\mbox{all pairs without $\mu_1,\mu_2$}) ...,0,0,...}
\eeqa
There are $N-2$ zeroes in the end, corresponding to elements of the Cartan subalgebra other than $T^3$.
Let us check the size of the matrix (\ref{DiagAdj}):
$$3 \,+\, 4 (N-2)\, +\, (N-2)(N-3)\, +\, (N-2) = N^2-1$$ as it should be
for the adjoint representation of $SU(N)$.

This constant background is exactly equivalent to twisting the boundary conditions.
So we have to sum logarithms of determinants for one adjoint-representation $SU(2)$ dyon, $2 (N-2)$ differently
twisted fundamental-representation $SU(2)$ dyons and
a determinant for $(N-2)(N-3)$ different constant $A_4$ field eigenvalues.
It would be interesting to check that this is asymptotically the
same as the adjoint determinant for the far region that we will calculate below.

In \cite{GSFermSUN} we proved that the twisting of boundary conditions for the fundamental
representation determinant
results in shifting of the argument of $P$, where $P$ is the standard perturbative potential
\beq \la{defP}
P(\v) = \left.\frac{1}{3(2\pi)^2}\v^2(2\pi -\v)^2\right|_{{\rm  mod}\; 2\pi }.
\eeq
%Also there are $N$ types of different KvBLL solutions with holonomies,
%that differ by the element of the $SU(N)$ group center (we label them by $k=1..N$). This $k$ also results in shifting the
%argument of $P$, as discussed in \cite{GSFErmSUN}.
% There is no ``background'' added to the triplet representation, since the unit matrix is zero in the adjoint
%representation.

Now we can just write the result (for the first dyon, for simplicity),
as a sum of the formula taken from \cite{DGPS} for the triplet ($SU(2)$ adjoint representation):
\beqa
&&\d_{\P}\log\Det(-D^2)_{\rm near\; dyon}
\nn \\&& =\d_\P \left(\tilde c_{dyon}\nu_1-\frac{8}{3} \nu_1 \log(\nu_1)+ \log \nu_m + \int^R P\left(2 \pi \nu_1-\frac{1}{r}\right) 4 \pi r^2 dr\right)
\label{intM}
\eeqa
 with $2(N-2)$ formulas from \cite{GSFermSUN} for the $SU(2)$ dyon with twisted boundary conditions
 (``twist'' is a corresponding
  matrix element of \eq{DiagAdj})
  \footnote{Note that to get the normal $SU(2)$ dyon we have to center the interval $(\mu_1,\mu_2)$ at zero, because
  in $SU(2)$ we obviously have the zero-trace condition $\mu_1+\mu_2 = 0$}:
\beqa\label{intM2}
&& \d_{\P}\log\Det(-\nabla^2)_{\rm near\; dyon} =\sum_{i=1}^{2(N-2)} \d_\P \left\{\hat c_{dyon}\nu_1-\frac{\log(\nu_1)}{6}\nu_1 \nn \right.
\\  && \left. +
\int^R \frac12 \left[P\left(2 \pi (\mu_1 -(\mu_1+\mu_2)/2) + \frac{1}{2 r}  -i (A_{const}^{adj})_{2 i+1} \right) \right.\right.\nn
\\ && \left. \left.  + P\left(2 \pi (\mu_{2} -(\mu_1+\mu_2)/2) - \frac{1}{2 r}
-i (A_{const}^{adj})_{2 i+2} \right) \right] 4 \pi r^2 dr\right\}
\eeqa
 and with a constant-field determinant for the remaining matrix elements of $ -\!D^2 $ ( i.e. twists without background field)
\beqa \nn
\d_{\P}\log\Det(-D^2)_{\rm near\; dyon}^{\rm const}&=&\!\!\!\!\sum_{i=1}^{(N-2)(N-3)} \!\!\!
\d_\P \int_0^R \frac12 P\left(-i (A_{const}^{adj})_{2+4 (N-2)+i} \right) 4 \pi r^2 dr.
\eeqa
Totally we get:
\beqa
\d_{\P}\log\Det(-D^2)_{\rm near\;dyon} &=& \d_{\P} \left(c_{dyon}\nu_m
- \left(\frac{8}{3}+\frac {N-2}3  \right)   \log(\nu_m) \nu_m  + \log \nu_m \right) \nn \\
+(R {\rm - terms})
\eeqa
And it is easy to check explicitly that the ``R-terms'' exactly match the asymptotics of far-from-dyons domain
(see the Section \ref{secFar})
\beq
\la{Rterms}
 (R{\rm -terms}) = \lim_{r_{i\neq 1} \to \infty} \int^R  \sum_{i>j}   \delta P\left[2\pi(\mu_i-\mu_j)+\frac{1}{2 r_i}-\frac{1}{2 r_{i-1}}-\left(\frac{1}{2 r_j}-\frac{1}{2 r_{j-1}}\right)\right] 4 \pi r_1^2 dr_1
\eeq

% Specializing to $\P = \mu_n$ and using $\nu_i \equiv \mu_{i+1} - \mu_i$ we get
%\beqa
%&&\d_{\mu_n}\log\Det(-D^2)_{\rm near\;m^{th}\;dyon} \nn = \\&& \d_{\mu_n} \left(c_{dyon}\nu_m
%- \left(\frac{8}{3}+\frac {N-2}3  \right)   \log(\nu_m R) \nu_m \right) +
%\frac 1 {\nu_m} \delta_{n,m+1} - \frac 1 {\nu_m} \delta_{n,m}
%+(R {\rm-\; terms}).
%\eeqa

%This result is a sum of $SU(2)$ adjoint determinant and $N-2$ double fundamental determinants with non-trivial
%boundary conditions. It was proved that boundary conditions only shift the argument of the potential (this
%shift does not alter the $\log R$ term).
 So we conclude that the ``R - terms'' are trivial and  exactly match the contributions from the outer region,
as it should be, of course, since the result cannot depend
on the radius of the auxiliary balls that we have chosen.

\section{Calculation of the currents for outer domain} \la{AppA}

\subsection{Singular current}
  The contribution of the singular part of the propagator to the variation of the determinant is 4 times the fundamental
representation result \cite{BC, GPY} , if we write this variation in terms of fundamental representation.
So we just take our old result from \cite{GSFermSUN} (that formula was not written there explicitly, it was in our
intermediate computations). For the component ${J^\s}^i$ it is quite natural to introduce bipolar spatial coordinates
with unit repers $\hat r_i = \frac{r_i}{|r|}$ , $\hat s_i = \frac{r_{i-1}}{|r|}$ ,
$n_\phi = \frac{\hat r \times \hat s}{|\hat r \times \hat s|}$.  In these coordinates the current (already multiplied by 4) is

\beqa
\la{sing_Curent}
 {J^\s}_4^i &= &-\frac{i \left(r_i^3-s_i^3\right)}{12 \pi ^2 r_i^3 s_i^3} \\
 {J^\s}_\phi^i &=& -\frac{i \left(r_i+s_i\right) \sqrt{-\left(\varrho_i-r_i-s_i\right) \left(d+r_i-s_i\right)
   \left(\varrho_i-r_i+s_i\right) \left(\varrho_i+r_i+s_i\right)}}{4 \pi ^2 r_i^2 s_i^2
   \left(\varrho_i+r_i+s_i\right)^2}\\
 {J^\s}_{\hat r_i}^i  &=& 0 \;\;\;;\;\;\; {J^\s}_{\hat s_i}^i  = 0
\eeqa
 We also remind the notations: $r_i=x-y_i$ is a vector from the $i$-th dyon center to the current point,
$s_i \equiv r_{i-1}$; and $\varrho_i = |y_i-y_{i-1}|$ is a distance between these two dyons. Also standard
``circular rule'' $r_{N+1} \equiv r_1$ is implied.

\subsection{M-term current}
  Let us prove that the contribution to the current from the M-term of adjoint propagator is zero with exponential
precision (i.e. it decays exponentially out of the dyon cores).
As was shown in \cite{DGPS}, when making the propagator periodic the M-term simplifies to
\beq
{\cal G}^{\m \, ab}(x,y) \equiv \frac{1}{8\pi^2}\int_{-1/2}^{1/2}dz dz'  M(z,z')\,
\Tr\left(v^{2\dagger}(x,z) v^2(x,z)\tau^a\right)\,
\Tr\left(v^{2\dagger}(y,z') v^2(y,z')\tau^b\right),
\label{MtermFinal}
\eeq
since the property $
v(y^n,z)= e^{2\pi i n z} v(y,z)
$ used to derive that result, still holds for the $SU(N)$ ADHMN construction.
Here $M(z_1,z_2,z,z)=\delta(z_1-z_2)M(z_1,z)$.

First of all we note
that only the lower components of $v$ are left and only the Cartan (diagonal)
components are nonzero: From \eq{vFar} we see that for each $m$ the function $s_m(z)$ and hence
$v^2_m(z)$ is peaked near $z=\mu_m$ and exponentially decays away from this point. So $v^{2 \dag}_{m }(x,z) v^2_n(x,z)
\sim \delta_{mn}$ with exponential precision. This leads us to conclusion that
\beq
{\cal G}^{\m\,ab}(x,y) \propto \delta^{a\in Cartan}\delta^{b \in Cartan},\;\;\;\;\;
{\cal G}^{\m\,ab}(x,y)={\cal G}^{\m\, ab}(y,x) \; .
\eeq
The second equation means that the terms with derivatives in the expression
for the current \ur{defJ} cancel each other. It follows from the first one that
the adjoint action of $A$ on ${\cal G}^m$ gives zero since both lie approximately in the
Cartan subalgebra.  Therefore we conclude that
\beq
J^\m_\mu\simeq 0 \;.
\eeq

\subsection{Regular current}
 The adjoint-representation regular current is
\beq
 J^{ab} = D_x^{ac} {\cal G}_{cb}(x,y) + {\cal G}_{ac}(x,y) D_y^{cb}
 \label{jgeneral}
\eeq
where $a,b,c = 1..N^2-1$ and we take the regular part of the propagator:
\beq
({\cal G}^{\r})^{ab}(x,y)\equiv\sum_{n\neq 0}
\frac{4}{8\pi^2(x-y_n)^2}\Tr\left[t^a \<v(x)|v(y_n)\>t^b\<v(y_n)|v(x)\>\right],
\qquad y_n^i=y^i - \delta^{i 4} n,
\label{Gr}
\eeq
It is possible to rewrite these formulae in the fundamental notations and evaluate explicitly.
Some details of the calculation together with a short review of ADHM construction are presented below.
We denote the adjoint indices by $a,b,c = 1..N^2-1$ and fundamental indices by
$i,j,k,l,m,n = 1..N$.

First we represent the covariant derivatives in the fundamental representation.
With the help of the identities
%(see for example \cite{Zar})
\def\dl{\overrightarrow{D}_{\mu}}
\def\dr{\overleftarrow{D}_{\mu}}

\begin{eqnarray}\label{id}
 \dl^{ad}\tr
 \left(t^dAt^bB\right)=\tr\left[t^a(\dl A)t^bB-
 t^aAt^b(B\dr)\right], \nonumber\\
 \tr
 \left(t^aAt^dB\right)\dr^{db}=\tr\left[t^a(A\dr)t^bB-
 t^aAt^b(\dl B)\right],
 \end{eqnarray}

 one gets for $J^{ab}$ the obvious four terms plus
$\delta_{\mu 4} \frac{2 \tr[t^a v_x^\dag v_y t^b v_y^\dag v_x]}{\pi^2 n^3}$
from the derivative acting on the denominator.

All the terms in the adjoint current have the form $\tr[t^a B t^b C]$.
The variation of the determinant has the form $-\delta A^c T^c_{a b} J^{a b}$.
To write it in the fundamental representation we use the identities:
\beq
 A^c T^c_{ab} = 2 \tr(t^b [t^a,A])
\eeq where $A=A^i t^i$
and
\beq
 t^a_{i j} t^a_{k l} = 1/2 (\delta_{il}\delta_{jk} - 1/N \delta_{ij} \delta_{kl})
\eeq
The  Hermitian generators $t^a$ are normalized as $\tr(t^a t^b) = 1/2$ (for $SU(2)$ these are $\tau^a/2$).
  We get:
\beq
 \delta A^c T^c_{a b} \tr(t^a B t^b C) = \frac12 \tr(B)\; \tr(\delta A \;C) - \frac12  \tr (C) \; \tr (\delta A\; B)
\label{curr1}
\eeq
  So in terms of the {\it fundamental} indices ($i,j = 1 ... N$) we get for the current $J^{ij}$ (that is to be coupled to
$A$ in the {\it fundamental} representation to get the variation of determinant)
\beqa
({J}^{\r}_\mu)^{ij}(x) &=& \sum_{n\neq 0,\{B,C\}}
\frac{1}{4\pi^2(x-y_n)^2} \left(\tr(B)\; C^{ij} -  \tr (C) \;  B^{ij} \right) \nn \\&&-
\delta_{\mu 4} \sum_{n\neq 0}
\frac{1}{\pi^2(x-y_n)^3} \left(\tr(E)\; F^{ij} -  \tr (F) \;  E^{ij} \right)
\label{Jr1}
\eeqa
Here
\def\vx{{v_x}}
\def\vy{{v_{y_n}}}
\def\vxd{{v_x^\dag}}
\def\vyd{{v_{y_n}^\dag}}
\beq
E = \vxd \vy \equiv b\;\; ; \;\; F = \vyd \vx \equiv b^\dag
\eeq
and we put $y=x$ (so that now $ y_n^i=x^i - \delta^{i 4} n$ ) according to \eq{defJ}.
The set $\{(B,C)\}$ consists of 4 pairs taken from \eq{id} and \eq{jgeneral}:
\beq
\{(B,C)\}=\{(D_x \vxd \vy, \vyd \vx) ,\; (\vxd \vy, D_x \vyd \vx),
\; (\vxd\vy D_y, \vyd \vx), \; (\vxd \vy, \vyd \vx D_y)\}
\label{list}
\eeq
 Since the current and the field are approximately diagonal in the fundamental representation,
 we consider the diagonal components $(J^\r_\mu)^{i } \equiv (J^\r_\mu)^{i i}.$
%Consider $j$-th diagonal matrix element under the first trace in \eq{curr1}  and $i$-th component
%of the second trace in \eq{curr1} (i.e. coupling with $i$-th diagonal in fundamental repr.
%component of $\delta A$).
  In these notations the contribution to equation \eq{Jr1} can be rewritten as
\beq
({{J}^{\r}}_\mu)^i(x) =\sum_{j=1}^N \sum_{n\neq 0} \left( \sum_{\{(B,C)\}}
\frac{1}{4\pi^2(x-y_n)^2} \left(B_j\; C_i -  C_j \;  B_i \right)-
\frac{\delta_{\mu 4}}{\pi^2(x-y_n)^3} \left(E_j\; F_i -  F_j \;  E_i \right) \right)
\label{Jr2}
\eeq

%So we wish to compare the result with the expected one:
%\beq
%({J_i}^{})(x,y) =^{\!\!\! ?}  \sum_j \frac12 P'\left[2\pi(\mu_i-\mu_j)+\frac{1}{2 r_i}-\frac{1}{2 s_i}-\left(\frac{1}{2 r_j}-\frac{1}{2 s_j}\right)\right]
%\eeq

From \eq{list} and \eq{Jr2} we get 8+2 terms in the resulting contribution to the current. Now in order to calculate explicitly
these $B,C,E,F$ we need ADHMN construction. A brief review and a calculation follows.

\subsection{Expressions of the ADHM construction} \la{secRegCurrent}
The basic object in the ADHMN construction~\cite{ADHM,Nahm80} is the
$(2+N)\times 2$ matrix $\Delta$ linear in the space-time variable $x$ and depending
on an additional compact variable $z$ belonging to the unit circle:
\beq
\Delta^K_\beta(z,x)=\left\{\bea{cl}
\lambda^m_\beta(z)&,\quad K=m,\quad 1\leq m\leq N,\\
(B(z)-x_\mu\sigma_\mu)^\alpha_\beta&,\quad K=N+\alpha,\quad 1\leq \alpha\leq 2,
\eea\right.
\la{Delta}\eeq
where $\alpha,\beta=1,2$ and $m=1,\dots,N$; $\sigma_\mu=(i\vec\sigma,1_2)$. As usual,
the superscripts number rows of a matrix and the subscripts number columns.
The functions $\lambda^m_\beta(z)$ forming a $N\times 2$ matrix carry information about
color orientations of the constituent dyons, encoded in the $N$ two-spinors $\zeta$:
\beq
\lambda^m_\beta(z)=\delta(z-\mu_m)\zeta^m_\beta.
\la{deflambda}\eeq
The quantities $\zeta^m_\beta$ transform as contravariant spinors of the gauge group
$SU(N)$ but as covariant spinors of the spatial $SU(2)$ group.
The $2\times 2$ matrix $B$ is a differential operator in $z$ and depends on the positions
of the dyons in the $3d$ space $\vec y_m$ and the overall position in time $\xi_4=x_4$:
\beq
B^\alpha_\beta(z)=\frac{\delta^\alpha_\beta\d_z}{2\pi i}+\frac{\hat A^\alpha_\beta(z)}{2\pi i}
\la{defB}
\eeq
with
\beq
\hat A(z)=A_\mu\sigma_\mu,\qquad \vec A(z)=2\pi i\, \vec y_m(z),\qquad A_4=2\pi i\,\xi_4,
\la{eq1}\eeq
where inside the interval $\mu_m\leq z\leq \mu_{m+1}$, $\vec y(z)=\vec y_m$ is the position
of the $m^{\rm th}$ dyon with inverse size $\nu_m\equiv \mu_{m+1}-\mu_m$.

One has to find $N$ quantities $v^K_n(x),\,n=1...N,$
\beq
\la{defv12}
v^K_n(x)=\left\{\bea{cl}v^{1m}_n(x)&,\quad K=m,\quad 1\leq m\leq N,\\
v^{2\alpha}_n(z,x)&,\quad K=N+\alpha,\quad 1\leq \alpha\leq 2,
\eea\right.
\eeq
which are normalized independent solutions of the differential equation
\beq
{\lambda^\dag}^\alpha_m(z) v^{1m}_n+[B^\dag(z)-x_\mu\sigma^\dag_\mu]^\alpha_\beta v^{2\beta}_n(z,x)=0,\qquad
v^{\dag 1m}_l\,v^{1l}_n+\int_{-1/2}^{1/2}dz\,v^{\dag 2m}_\alpha\,v^{2\alpha}_n=\delta^m_n\,,
\eeq
or, in short hand notations,
\beq
\la{zerov}
\Delta^\dag v=0,\qquad v^\dag v=1_N.
\eeq
Note that only the lower component $v^2$ depends on $z$.

Expressing $v$ as
\beq
v(x)=\left(\bea{c}-1_n\\u(x)\eea\right)\phi^{-1/2},\;\;\;\;\;u(x)=(B^\dag-x^\dag)^{-1}\lambda^\dag
\eeq
let us find $u(z,x)$ -- the main object of ADHM construction.
It is the solution to the equation
\beq
(B^\dag-x^\dag)u=\lambda^\dag,\;\;\;\;\;B^\dag-x^\dag=\frac{\d_z}{2\pi i}-r^\dag(z)
\eeq

Define the Green's functions:
\beq
f=(\Delta^\dag\Delta)^{-1},\;\;\;\;\;G=((B-x)^\dag(B-x))^{-1},\;\;\;\;\;\phi_{ij}(x)=\delta_{ij}+\lambda_{i}^{\alpha} G^{\alpha\beta} {\lambda^\dag}^{\alpha}_j
\eeq
where $(i=1\dots N)$, $(\alpha,\beta=1,2)$.  One can note that
\beq
f=(G^{-1}+\lambda^\dag\lambda)^{-1}=G-G\lambda^\dag_i\phi_{ij}^{-1}\lambda G
\la{ftG}
\eeq
acting on (\ref{ftG}) with $\lambda^\dag$ on the right yields
\beq
G\lambda^\dag_j=f\lambda^\dag_i\phi_{ij}
\la{ftG2}
\eeq

The Green's function is expressed as follows:
\beq
f(z,z')=s_m(z) f_{mn}s^{\dag}_n(z')+2\pi s(z,z')\delta_{[z][z']}
\la{fzz}\eeq

\beq
f_{mn}={F^{-1}}_{mn}
\eeq

The functions appearing
in \eq{fzz} are
\beqa
\nn s_m(z)&\!\!\!=&e^{2\pi i x_0(z-\mu_m)}\frac{\sinh[2\pi r_m(\mu_{m+1}-z)]}{\sinh(2\pi r_m\nu_m)}\delta_{m[z]}
+e^{2\pi i x_0(z-\mu_{m})}\frac{\sinh[2\pi r_{m-1}(z-\mu_{m-1})]}
{\sinh(2\pi r_{m-1}\nu_{m-1})}\delta_{m,[z]+1},\\
\nn s(z,z')&\!\!\!=&e^{2\pi i x_0(z-z')}\frac{\sinh\!\left(2\pi r_{[z]}
(\min\{z,z'\}-\mu_{[z]})\right)\sinh\!\left(2\pi r_{[z]}
(\mu_{[z]+1}-\max\{z,z'\})\right)}{r_{[z]}\sinh\!\left(2\pi r_{[z]}\nu_{[z]}\right)}.
\eeqa

\beq
u_i=(B-x)f\lambda_j^\dag\phi_{ji}=\left(\frac{\d_z}{2\pi i}-r_\mu\sigma_\mu\right)s^f_k(z)f_{kj}\zeta^\dag_j\phi_{ji}
\eeq

And the convenient notation is $r_i=x-y_i$ is a vector from the $i$-th dyon to the current point.
First we note that $F_{ij}$ and $\phi_{ij}$ are diagonal matrices with exponential precision
\beq
f_{ij}\simeq 2\pi \delta_{ij}(r_i+r_{i-1}+\vrho_i)^{-1}
\eeq

\beq
\phi_{ij}\simeq\delta_{ij}\frac{r_i+r_{i-1}+\vrho_i}{r_i+r_{i-1}-\vrho_i}
\eeq

To pass to the periodical gauge we multiply $v(x)$ from the right by
$g_{ij}=\delta_{ij} e^{2 \pi i \mu_i x_0}$.
%Let us denote $(B-x) = D_x$.
Totally within the exponential precision we get for $v$:
\beq
\la{vFar}
 v_i(x,z)= \left(\bea{c}-\delta_{ij} \phi_{ii}^{-1/2} e^{2 \pi i \mu_i x_0}\\
 (B-x) s_i(z) f_{ii} \zeta_i^\dag \phi_{ii}^{1/2} e^{2 \pi i \mu_i x_0}\eea\right) \;\;\;\mbox{no index summations}
\eeq

Consider the covariant derivative of $v_i(x)$ in the periodical gauge (integration over $z$ is assumed):
\beqa
D_\mu\<v(x)|&\!\!\!= &\!\!\!\d_\mu\<v|-\d_\mu\<v|v\>\<v|= \d_\mu\<v|(1-|v\>\<v|)\\
\nn &\!\!\!= &\!\!\!\d_\mu\<v|\Delta f\Delta^\dag = -\<v|\d_\mu\Delta f\Delta^\dag\\
\nn &\!\!\!= &\!\!\!-\<v|\cB \sigma_\mu f\Delta^\dag  =
\left(v^{2 \dag} \sigma_\mu f \lambda^\dag,v^{2\dag} \sigma_\mu f (B-x)^\dag \right)\\
&\!\!\!=&
\nn  f_{ii} \zeta_i \phi_{ii}^{1/2} e^{-2 \pi i \mu_i x_0} (B-x)^\dag s_i^\dag(z) \sigma_\mu \left(
f \lambda^\dag, f (B-x)^\dag\right)\\
&\!\!\!=&
\nn f_{ii} \zeta_i \phi_{ii}^{1/2} e^{-2 \pi i \mu_i x_0} (B-x)^\dag s_i^\dag(z) \sigma_\mu \left(
s_i(z) f_{ii} \zeta^\dag_i, \left(s_i(z) f_{ii} s_i^\dag(z') + 2 \pi s(z,z')\right) (B-x)^\dag\right)
 \;.
\eeqa

\beq
 |v(x)\> D_\mu = - (D_\mu\<v(x))^\dag
\eeq
\subsection{Formula for the regular current}
 Let us denote $D=(B-x)$. We will in a moment express  \eq{list}
 through $c$ and $b$, defined as
\beqa
 c&=&D_{\mu} v_x^\dag v_y \\ \nn
 &=& - \phi n f_{ii}^2 \zeta_i (Ds(z))^\dag \sigma_\mu (s(z) f s^\dag(z') + 2 \pi s(z,z'))
 e^{-2 \pi i n (z'-\mu_i)} Ds(z') \zeta_i^\dag e^{- 2 \pi i \mu_i n}\\ \nn
 &=& - \phi n f_{ii}^2 \zeta_i (\tilde D \tilde s(z))^\dag \sigma_\mu (\tilde s(z) f \tilde s^\dag(z') + 2 \pi \tilde {s}(z,z'))
 \tilde D \tilde s(z') \zeta_i^\dag e^{-2 \pi i n z'}
\eeqa
here $\,\tilde{}\,$ means that the time dependence is separated (so that $\tilde s$ is time-independent)
and integration over $z$ and $z'$ is assumed. To derive this we used
\beq
D^\dag_x  v_y^2 = (D^\dag_y - n) v_y^2 = - \lambda v_y^1 -n v_y^2
\eeq
following from \eq{zerov} and noticed that the first term cancels with the scalar product of upper components.
For $b$ we easily get
\beqa
 b=v_x^\dag v_y = \exp(-2 \pi i \mu_i n)/\phi_i + \phi_i f_{ii}^2 \zeta_i ((D s)^\dag e^{- 2 \pi i n z} Ds) \zeta_i^\dag.
 \eeqa
We also need the following formulas:
\beqa
  v_x^\dag v_y D_\mu^{(y)} &=& -c^\dag_{n\to-n}\\
  v_y^\dag v_x &=& b^\dag \\
  b_{n \to -n} &=& b^\dag \\
 v_y^\dag v_x D_{\mu}^{(y)} &=& - D_{\mu}^{(y)} v_y^\dag v_x + 2 A_\mu v_y^\dag v_x= -c_{n\to -n} + 2 A b^\dag \\
 D_\mu^{(x)} v_y^\dag v_x &=& -v_y^\dag  v_x D_\mu^{(x)} + 2 A v_y^\dag v_x = c^\dag + 2 A b^\dag
\eeqa

In the total sum changing $n \to -n$ does not affect the expression since we divide by $n^2$, so
we can make $n\to -n$ in the whole expression by conjugating $b$ and then drop $n \to -n$.

Then the set of $\{(B,C)\}$ in \eq{list} becomes
\beq
\{(B,C)\}=\{(c,b^\dag), \;(b,c^\dag+ 2 A b^\dag), \; (-c^\dag_{n\to -n},b^\dag),\; (b,-c_{n\to-n}+2 A b^\dag)\}
\label{list1}
\eeq

Totally we get for the current (\ref{Jr2})  (the index $\mu$ is hidden in $c$):
\beq
 ({{J}^{\r}}_\mu)^i(x) = \sum_{n\neq0 ; j=1..N} \frac{2 c_j b_i^\dag - 2 c_j^\dag b_i - 4 A_j b_i b_j^\dag }{4 \pi n^2}
 -  \delta_{\mu 4} \frac{b_j b_i^\dag}{\pi^2 n^3} -
 (i\leftrightarrow j)
\eeq

There are only terms with $(i) \neq (j)$ in the current that we are calculating.

%To compare with $SU(2)$ calculations it is convenient to introduce:
%\beq
% d = \d_\mu \vxd \vy = c - A \,b
%\eeq
% Really, it's much simpler to evaluate $d$ directly.

%Then the set of $\{(B,C)\}$ is
%\beq
%\{(B,C)\}=\{(d+A\, b , b^\dag), \;(b, d^\dag+ A  \, b^\dag), \; (-d^\dag + A \,b , b^\dag),\; (b,-d+A \,b^\dag)\}
%\label{list2}
%\eeq

%Then we get for the current (the index $\mu$ is hidden in $d$):
%\beqa
% ({{J}^{\r}}_\mu)^i(x) &=& \sum_{n\neq0 ; j=1..N} \frac{2 d_j b_i^\dag - 2 d_j^\dag b_i +(
% 2 A_{\mu j} b_j b_i^\dag + 2 A_{\mu j} b_j^\dag b_i) - 4 A_{\mu j} b_i b_j^\dag }{4 \pi n^2}
% -  \delta_{\mu4} \frac{b_j b_i^\dag}{\pi^2 n^3} -
% (i\leftrightarrow j) \\
% &=& \sum_{n\neq0 ; j=1..N} \frac{2 d_j b_i^\dag - 2 d_j^\dag b_i }{4 \pi n^2}
% -  \delta_{\mu4} \frac{b_j b_i^\dag}{\pi^2 n^3} -
% (i\leftrightarrow j)
%\eeqa
%in the last line we used that $b_i b_j^\dag = \left. b_j b_i^\dag \right|_{n \to -n}$.

Now by these explicit formulas the current can be evaluated by performing integrals over $z$,$z'$ and summing over $n$. The
reference formulas for summation can be found e.g. in \cite{DGPS}. We used {\rm Mathematica} for these calculations.
Below  the result is presented.

Just for illustration and consistency check we write first for the $SU(2)$.
For $SU(2)$ the ADHMN data is taken to be $\mu_1 = -\omega, \;\; \mu_2= \omega \;,\; \nu_1 = 2 \omega, \;\; \nu_2 =  1-2 \omega $.
So we get for the current (here we write the diagonal matrix, which couples to the gauge field in
the fundamental representation)
\beq
 (J^r_4)^1 = -(J^r_4)^{2} = \left( i P'\left(2 \pi (\mu_1 - \mu_2) + \frac1r - \frac1s \right)+\frac{i}{12 \pi^2} \left( \frac1{s^3} - \frac1{r^3} \right)\right)
\eeq
As usual, the current is expressed as a derivative of perturbative potential $P$.
The second term in this expression cancels exactly with the contribution of ``singular'' current.
Totally, adding the singular current, we get for the ``far region'' contribution to variation:

\beqa
\delta \log \Det (\!-\!D^2) _{far} &=& \int_{far}\delta (A^{fund}_{41} - A^{fund}_{42}) P'\left(2 \pi (\mu_1 - \mu_2) + \frac1r - \frac1s\right)
\nn \\ &=& \int_{far} \delta  P\left(2 \pi (\mu_1 - \mu_2) + \frac1r - \frac1s \right)
\eeqa
 we remind the  convenient notation: $r_i=x-y_i$ is a vector from the $i$-th dyon to the current point,
$s_i \equiv r_{i-1}$ and a standard ``circular rule'' $r_{N+1} \equiv r_1$.
For $SU(2)$ we set $r=r_1$, $s=s_1=r_2$.

  For the $SU(N)$ case we get the total result:
\beqa
\la{result_far}
 &&\delta \log \Det (\!-\!D^2)_{far}  \nn
 \\ &=& \int_{far} \sum_{i,j =1}^N \delta\left(2 \pi \mu_i + \frac{1}{2 r_i}-\frac{1}{2 s_i}\right)
 \nn P'\left[2\pi(\mu_i-\mu_j)+\frac{1}{2 r_i}-\frac{1}{2 s_i}-\left(\frac{1}{2 r_j}-\frac{1}{2 s_j}\right)\right] {\rm sgn}(\mu_i - \mu_j)\\
&=& \int_{far} \sum_{i>j}   \delta P\left[2\pi(\mu_i-\mu_j)+\frac{1}{2 r_i}-\frac{1}{2 s_i}-\left(\frac{1}{2 r_j}-\frac{1}{2 s_j}\right)\right]
 \eeqa
 Note that the spatial part of the regular current cancelled exactly with the singular part of the current.
This proves the \eq{toint}.

%\section{The reduction to SU(2)}
%  Let us try to remove the $n$-th dyon ($\mu_n \to \mu_{n+1}$). It results in the following change of
%  the far contribution:
%\beqa
%\eeqa
%  Let us consider in more detail the $\log \Pi$ terms, which are new and obviously vanish for the
%  $SU(2)$ case.

\end{document}